# Verification of Peterson's Algorithm for Leader Election in a Unidirectional Asynchronous Ring Using NuSMV

Amin Ansari

Electrical Engineering and Computer Science Department
University of Michigan, Ann Arbor, MI
ansary@eecs.umich.edu

**Abstract.** The finite intrinsic nature of the most distributed algorithms gives us this ability to use model checking tools for verification of this type of algorithms. In this paper, I attempt to use NuSMV as a model checking tool for verifying necessary properties of Peterson's algorithm for leader election problem in a unidirectional asynchronous ring topology. Peterson's algorithm for an asynchronous ring supposes that each node in the ring has a unique ID and also a queue for dealing with storage problem. By considering that the queue can have any combination of values, a constructed model for a ring with only four nodes will have more than a billion states. Although it seems that model checking is not a feasible approach for this problem, I attempt to use several effective limiting assumptions for hiring formal model checking approach without losing the correct functionality of the Peterson's algorithm. These enforced limiting assumptions target the degree of freedom in the model checking process and significantly decrease the CPU time, memory usage and the total number of page faults. By deploying these limitations, the number of nodes can be increased from four to eight in the model checking process with NuSMV.

## 1 Introduction

Often, selecting a unique leader among different processes becomes necessary when these processes are working on a distributed system. For instance, broadcasting [2] and global synchronization on a distributed system are the cases that the leader election becomes applicable. In general, a distributed system can work on any network with an arbitrary topology. One of the most primitive and prevalent topologies for distributed processing is ring topology [6]. In a ring topology, each node only connects to its two adjacent neighboring nodes. These links can establish a bidirectional or a unidirectional communication between the two connected neighbor nodes in the ring. A ring topology allows the nodes to have a synchronous or an asynchronous communication. In a synchronous ring, each node can send a message to its neighbors only in specific predefined synchronous time events (clock), while in an asynchronous ring each node can send its message in any random time that it becomes ready for sending that to the other neighboring nodes in the ring.

There are several algorithms for solving the leader election problem in different network topologies [3], [7], [8], [9], [14], [15]. Most of the previous works on leader election problem focused on a synchronous ring. By the way, synchronization of all nodes in a large scale problem or in a realistic huge ad-hoc application is not a worthwhile affordable approach. Thus, it is more realistic to work on asynchronous networks for dealing with huge and not well-defined topologies. The various proposed algorithms for solving the leader election problem have different properties based on the number of messages which they pass through the network and also the number of phases that each algorithm should run to select a leader. One of these leader election algorithms was presented by Peterson for determining the leader in a unidirectional asynchronous ring. Peterson's algorithm solves efficiently this problem in the best known order which only passes $O(n \log(n))$ message through the network [3].

This algorithm assumes that each node in our ring network has a unique identification number (UID). Peterson [3] and Dolev, Klawe, and Rodeh [9] independently extended the Franklin solution [13] for a bidirectional ring to a unidirectional ring [8]. Peterson's algorithm supposes that there is a FIFO queue in each node of the network and this allows each node to send leader election related messages to its neighboring nodes at any time. There are several previous works on the verification of other leader election algorithms for synchronous and asynchronous ring networks by using various methods and tools. Some of the new methods for leader election in a ring use probability theory and probabilistic models such as Itai-Rodeh algorithm [7], [8]. It means that the algorithm does not always return the leader in the ring; although, it gradually increase the probability of the leader election approximately to one in the long-term while the program is running.

Although there are previous works on the formal verification of the leader election problem for different networks [1], [4], [11], [12], I cannot find any published work on the formal verification of the Peterson's algorithm, using model checking with symbolic model verifier (SMV) tools. Because of the tremendous number of different combination of values in the queue of each node; it seems that the verification of the Peterson's algorithm is not feasible for the considerable number of nodes in a ring network. SMV as a verification tool works efficiently on any problem which can be formulated by a transition system and a finite reasonable number of states. For example, since the mutual exclusion algorithms for few processes do not have considerable number of states in their transition model, SMV and other symbolic model checkers are normally used for the formal verification of the mutual exclusion algorithms [10], [16], [17].

Formal verification methods also widely have been used to validate the coherence protocol of the SGI Origin2000[1] [21], [22], [23] and the Sun Scalable Shared-Memory Multiprocessor (S3.mp) [19], [20]. Eiriksson used the SMV [22] to verify an abstract model of a three node Origin2000 system which consists of two processors and an I/O unit. Pong, et. al. used the Mur$\varphi$ [19] formal verification system to verify an abstracted three node Sun Scalable Shared-Memory Multiprocessor (S3.mp) system. Each of these efforts needed to model the system at a high-level of abstraction (ignoring as much detail as possible) to make the verification tractable. Despite the unrealistically small verification model, these methods proved to be very valuable in extracting very subtle design flaws that would have been difficult, if not impossible, to detect using traditional simulation methods [18].

The main challenge through this paper is to reduce the state space of the Peterson's algorithm for the leader election problem, such that the model checking becomes feasible for larger rings. However, the different possible orders of the execution for the large number of processes makes the model checker to go through the all possible execution paths. Therefore, if the total number of states can be reduced significantly, model checking approach would not be still feasible for the situations where there are a lot of processes in the model. Nevertheless, it is still possible to model many applications with only a small set of processes.

In the rest of this paper, first, I describe the general case of the Peterson's algorithm and a suitable implementation of the algorithm for verification purpose. After that, I demonstrate the two modified versions of the algorithm with the implemented limiting assumptions. In the next step, necessary expected properties of Peterson's algorithm in the model checking process are discussed. And the final part of the paper exhibits the NuSMV model checking results and a brief summarization of the work.

## 2  Model Checking in the SMV System

Model checking is a method for formally verifying finite-state concurrent systems. Specifications about the system are expressed as temporal logic formulas, and efficient symbolic algorithms are used to traverse the model defined by the system and check if the specification holds or not [25]. That means that instead of writing simulation vectors or a simulation test bench, you verify your design for all possible input sequences. While formal verification is often equated with equivalence checking, model checking is substantially more general. It allows you to verify that that your specifications are correct very early in the design process by building abstract system level models. Its use continues through the design refinement process, allowing you to verify that your RTL level design correctly implements your high level model [26]. SMV is a tool for checking that finite-state systems satisfy specifications given in computation tree logic (CTL). It used the Ordered Binary Decision Diagram (OBDD) bases symbolic model checking algorithm [24]. The primary purpose of the SMV input language is to describe the transition relation of a finite Kripke structure. The input file describes both the model and the specification. The states are defined by a collection of state variables, which may be of Boolean or scalar type. The transition relation of the Kripke structure is determined by a collection of parallel assignments. The semantics of assignment in SMV is similar to that of a single assignment data flow languages. A program can be viewed as a system of simultaneous equations, whose solutions determine the next state. When a set is assigned to a variable, the result is a non-deterministic choice among elements of the set. Non-deterministic choices are useful for describing systems which are not yet fully implemented or abstract models of complex protocols, where the value of some state variables cannot be completely determined [10].

---

[1] The machine that has an aggressive cache-coherent communication architecture

## 3   General Peterson's algorithm

Each node in the Peterson's algorithm can be in an active or a relay state. When the algorithm starts to work, all nodes will be initiated to be in the active mode. Each node has a virtual ID tag which is initiated to the UID of the node in the beginning of the execution.

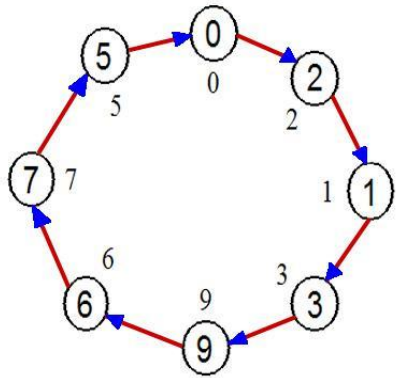

**Fig. 1.** An eight node unidirectional ring is shown just after the initialization step.

Figure 1 shows the state of an eight node unidirectional ring after the initialization step. The numbers in the circles present the UID and the numbers outside of the circles show the virtual ID tags. As it can be seen, just after initialization, all the virtual ID tags and UIDs are the same.

In the active mode, each node sends its current virtual ID tag to the next node and waits until it receives the first message from the neighboring node. After that, it compares the received message to its current virtual ID tag. If the ID matches, it will announce itself as the leader of the ring. Otherwise, each node passes the received message to the next neighboring node in the ring and waits to receive the next message. When each node receives the next message, it compares the first received message with the new received one and its virtual ID tag. If the first received message is greater than both of the second message and its virtual ID tag, it will update its virtual ID tag with the first received message. Otherwise, it will enter the relay mode. In relay mode, the node just passes the received message to the next neighboring node in the ring[2] without doing any further process on the received messages.

Figure 2 shows the first complete cycle of the Peterson's algorithm on the previously described ring. Since the Peterson's algorithm works on an asynchronous ring, it should be noticed that the figure 2 is just one possible case. For instance, consider the node with UID 5, since its previous node has the greater virtual ID tag than 5 and 6; the virtual ID tag of the node 5 will be updated to 7 and it still remains in the active mode. Conversely, consider the node with UID 0, since the virtual ID tag of its previous node is less than 7; the node 0 will enter the relay mode. In figure 2, the nodes which entered the relay mode were shaded with gray color.

---
[2] Store and Forwarding

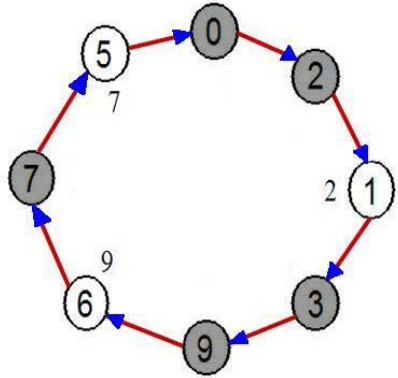

**Fig. 2.** An eight node unidirectional ring is shown after the first complete cycle of the execution of the Peterson's algorithm.

In other words, the Peterson's algorithm uses two modes for each node and it also uses message hopping for solving the leader election problem. Each message passes through two active nodes in the ring without concerning about the relay nodes. It means that the algorithm supposes that the relay nodes actually do not exist in the ring. Through this process, some nodes remain active and some others enter the relay mode. And the last remaining active node will become the leader of the ring. After that, the last active node will send the proper message through the network to inform other nodes about the new selected leader of the ring. Algorithm 1 shows the high level description of each node's behavior in the ring network.

**Algorithm 1** General Peterson's algorithm

```
VirtualID = UID;
Mode = Active;
while (TRUE) {
      if (Mode == Relay) {
            tempid = receive();
            send(tempid);
      } else {
            send(VirtualID);
            id2  = receive();
            if (VirtualID == id2)
                  announce("I'm the leader of the ring.");
            else {
                  send(id2);
                  id3 = receive();
                  if (id2 > max{VirtualID,id3})
                        VirtualID = id2;
                  else
                        Mode = Relay;
            }
      }
}
```

## 4  Implementation of Algorithm

For implementation of the Peterson's algorithm, first of all, **send** and **receive** functions should be replaced with suitable codes which can be used through the model checking process. In this implementation, I suppose that each node has a buffer and the previous node in the network has access to the local buffer of its neighboring node. And each buffer has two pointers. One of these pointers is an external pointer which can be used by the previous node to figure out where it should put the next element. The other pointer is an internal pointer which is used by the node to read from its local buffer. Furthermore, I add a state variable for keeping track of the flow model of the program in NuSMV. When the leader of the network is eventually elected, the state variable becomes equal to an arbitrary unique (LEAD_VAL) value for indicating that event. For sending data, there is no need to wait and the data can be added to the buffer of the next node and its external pointer is shifted by one. For receiving data, each node should go through a while loop until its local pointer becomes unequal to its external pointer. It means that the previous node sent a message. Model checking of this algorithm with more than four nodes will take an unreasonable amount of time on an ordinary system. Therefore, I try to make the algorithm simpler for targeting more nodes in the model checking. Algorithm 2 shows the suitable implementation of the general Peterson's algorithm which can be easily translated to the SMV modeling language for performing symbolic model checking with NuSMV.

**Algorithm 2** Suitable Version for Model Checking

```
VirtualID = UID;
Mode = Active;
while (TRUE) {
        state = 0;
        if (Mode == Relay) {
                while (inpointer == mypointer);
                nextbuff[nextpointer] = mybuff[inpointer];
                inpointer = (inpointer + 1) % n;
                nextpointer = (nextpointer + 1) % n;
        } else {
                nextbuff[nextpointer] = current;
                nextpointer = (nextpointer + 1) % n;
                state = 1;
                while(inpointer == mypointer);
                uid2 = mybuff[inpointer];
                inpointer = (inpointer + 1) % n;
                state = 2;
                if (VirtualID == id2) {
                        state = LEAD_VAL;
                        // I'm the leader of the ring.
                } else {
                        nextbuff[nextpointer] = id2;
                        nextpointer = (nextpointer + 1) % n;
                        state = 3;
                        while(inpointer == mypointer);
                        id3 = mybuff[inpointer];
                        state = 4;
                        inpointer = (inpointer + 1) % n;
                        if (id2 > max{VirtualID,id3})
                                VirtualID = id2;
                        else
                                Mode = Relay;
                }
        }
}
```

## 5  Modified Version of Algorithm

For making verification feasible for a ring with more nodes, I try to add several limiting conditions to the general described version of the algorithm. This proposed version uses two limiting assumptions for each node. Firstly, it supposes that each node has buffer with size one for receiving data from the previous neighboring node in the ring. Next, it supposes that each node can check whether the next node's buffer is empty or not. In this version, each node has to wait in the while loop for sending data which is not the case in the general version of the algorithm. Although these changes add one new value to the set of possible values of the state variable, it eliminates a myriad of states which are generated by the various placements of the values in the buffer of each node. These two mentioned modifications would not make the verification results to become invalid for the general algorithm. As can be seen in algorithm 2, each node generates its outputs in a sequential manner. It means that there is no indeterminism in the order of the elements in the queue of the next node. This sequential generation of the date reveals that having only one buffer in the receiver will not cause that the verification result becomes invalid for the general version. The modified version allows performing the model checking for a ring with up to six nodes with the same system in a reasonable amount of time. Algorithm 3 shows the high level description of each node's behavior for this modified version.

**Algorithm 3** Modified Peterson's algorithm

```
VirtualID = UID;
Mode = Active;
while (TRUE) {
    state = 0;
        if (Mode == Relay) {
                while (myinput == Empty);
                VirtualID = myinput;
                myinput = Empty;
                state = 1;
                while (nextinput != Empty);
                nextinput = VirtualID;
        } else {
                while (nextinput != Empty);
                nextinput = VirtualID;
                state = 2;
                while(myinput == Empty);
                id2 = myinput;
                myinput = Empty;
                state = 3;
                if (VirtualID == id2) {
                        state = LEAD_VAL;
                        // I'm the leader of the ring.
                } else {
                        while (nextinput != Empty);
                        nextinput = id2;
                        state = 4;
                        while(myinput == Empty);
                        id3 = myinput;
                        myinput = Empty;
                        state = 5;
                        if (id2 > max{VirtualID,id3})
                                VirtualID = id2;
                        else
                                Mode = Relay;
                }
        }
    }
}
```

# 6   Extra Modified Version of Algorithm

I try to make the algorithm even simpler to perform model checking routine on a ring with more nodes. This extra modified version uses another limiting assumption compared to the previous version. This version supposes that when a node enters the relay mode, it does not need to use two different states for receiving and sending the messages. It combines these two actions and when the node's buffer becomes nonempty and also the next node's buffer becomes empty, just writes its buffer value into the next node's buffer without using two different phases for receiving the message and sending that. In other words, this version supposes that each relay node is a part of the wiring connection between the active nodes. However, I cannot use this assumption and reduce the number of states in the verification model for the active mode. Because, in the beginning of the algorithm's execution, it is possible that all nodes enter the wait state for receiving a message and no one sends its already prepared message. And this problem will result to a deadlock in the algorithm. Consequently, I can only use this assumption for the relay mode. As another improvement, I also delete one other value from the set of possible values for state variable. By hiring these two new changes in the modified version of the algorithm, NuSMV can be used to do the model checking for up to eight nodes on the same system. Algorithm 4 shows the high level description of each node's behavior for the extra modified version of the Peterson's algorithm.

**Algorithm 4** Extra Modified Peterson's algorithm

```
VirtualID = UID;
Mode = Active;
while (TRUE) {
    state = 0;
        if (Mode == Relay) {
                while (myinput == Empty);
                while (nextinput != Empty);
                nextinput = myinput;
                myinput = Empty;
        } else {
                while (nextinput != Empty);
                nextinput = VirtualID;
                state = 2;
                while(myinput == Empty);
                id2 = myinput;
                myinput = Empty;
                state = 3;
                if (VirtualID == id2) {
                        state = LEAD_VAL;
                        // I'm the leader of the ring.
                } else {
                        while (nextinput != Empty);
                        nextinput = id2;
                        state = 4;
                        while(myinput == Empty);
                        if (id2 > max{VirtualID,id3})
                                VirtualID = id2;
                        else
                                Mode = Relay;
                        myinput = Empty;
                }
        }
}
```

## 7  Verification Environment and Platform

For doing model checking in this project, I used NuSMV 2.4.3 as a symbolic model checker which is linked to the Zchaff SAT solver[3]. This model checking was done in Microsoft Windows XP with SP1. The machine which was used for model checking had an Intel Pentium 4 CPU with 2.0 GHz clock rate and 512 MB main memory.

## 8  Expected Properties Specification

For making the model checking easier and also making the definition of the specifications simpler, I use CTL for defining the expected specifications for the three different versions of the Peterson's algorithm which are proposed in this paper [5], [10]. For becoming assured that the leader election result would be correct, each of these mentioned versions of the algorithm should have the following three main properties. I express each of these three expected specifications in CTL:

1. Eventually a leader will be elected for the ring:
    *AF  (        (node0.state = LEAD_VAL) |*
    *         (node1.state = LEAD_VAL) |*

---

[3] I also used the ordinary version of NuSMV; with a rough comparison, it seems that the version which is linked with the Zchaff SAT solver works to some extent faster.

                    *(node2.state = LEAD_VAL) | ...*
     *)*

2. At most one node will be elected as the leader of the ring. In other words, in any configuration of the system there should not be a situation that two nodes announce themselves as the leader of the ring: (**NOT** should be returned for this CTL expression. It means that this CTL query should not hold.)
   *EF (     ((node0.state = LEAD_VAL) & (node1.state = LEAD_VAL)) |*
             *((node0.state = LEAD_VAL) & (node2.state = LEAD_VAL)) |*
             *((node1.state = LEAD_VAL) & (node2.state = LEAD_VAL)) | ...*
       *)*

3. When some node announces itself as the leader of the ring, its virtual ID tag should be equal to the greatest initialized UID value of all the nodes in the ring:
   *AG (     ((node0.state = LEAD_VAL) → (node0.VirtualID = MaxUID)) &*
             *((node1.state = LEAD_VAL) → (node1.VirtualID = MaxUID)) &*
             *((node2.state = LEAD_VAL) → (node2.VirtualID = MaxUID)) & ...*
       *)*

## 9 Model Checking Results

Model checking was done for the Peterson's algorithm with different nodes in the ring. As the implementation of the algorithms in SMV modeling language, I modeled each node of the ring as a separate process. Process structure gives this capability to the model checker to consider the different order of the execution of the algorithm for the nodes in the asynchronous ring. Also, I put a fairness condition for each of the nodes (processes):

    *FAIRNESS running*

Otherwise, it would be possible that a process never gets the execution turn or equivalently it would be possible that a node never sends any message in the network for its neighboring nodes. It means that even one cyclic phase of the algorithm could not be completed for all the nodes in the ring. As a result, the ring enters a deadlock situation and none of the nodes can proceed its normal job.

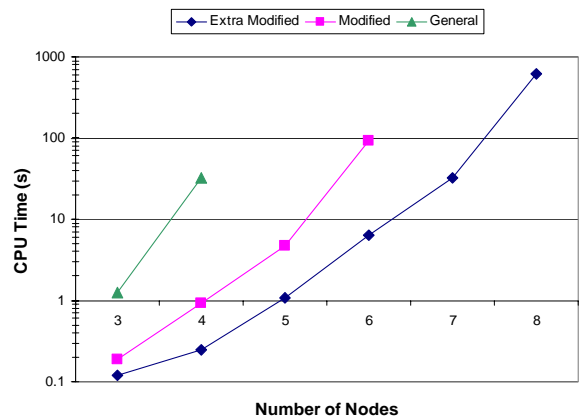

**Fig. 3.** Model checking execution time.

The CPU running time, virtual memory usage, memory size of the working set and the number of the page faults were extracted from Process Explorer which is an advanced process management utility. The results of performing model checking routine for the three mentioned properties on the different version of Peterson's algorithm can be seen in the figures 3 to 6. In Figure 3, execution time of model checker for each of the three different algorithms is shown based on the number of nodes that I have in the ring. Due to the fast growth in the model checking time, the vertical axis is shown in logarithmic scale. By looking to this figure, it can be concluded that for the general Peterson's algorithm increasing the number of nodes will result to the incredibly fast growth of the CPU time. And this growth becomes smoother when the modified version is used or even smoother when the extra modified version is used. It shows that the reasonable running time of the model checker on the extra modified version allows us to verify this version of the leader election algorithm for eight node ring in about 10 minutes.

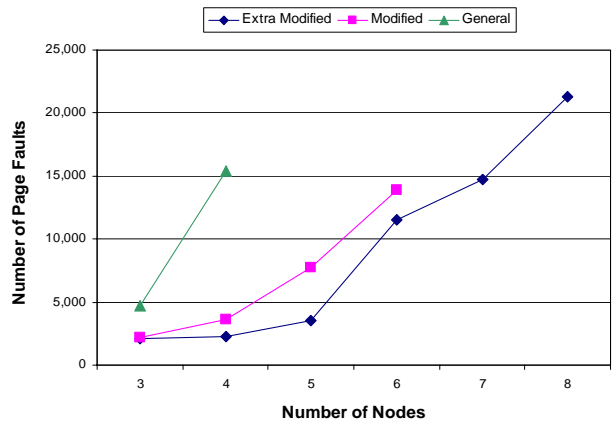

**Fig. 4.** Number of page faults in the model checking process for each case.

As it can be seen in the Figure 4, by adding nodes to the ring, the number of page faults will grow up approximately with a linear pace. Figure 4 shows that this linear function grows faster for the general version of the algorithm compared to the two other versions. Unjustified growth in the number of the page faults shows that the locality could not be used efficiently and the model checker mostly needs pages which are not in the memory at that time. This problem makes the operating system to bring the missed page from the hard disk to the memory for further usage by the model checker and also makes the verification time significantly longer.

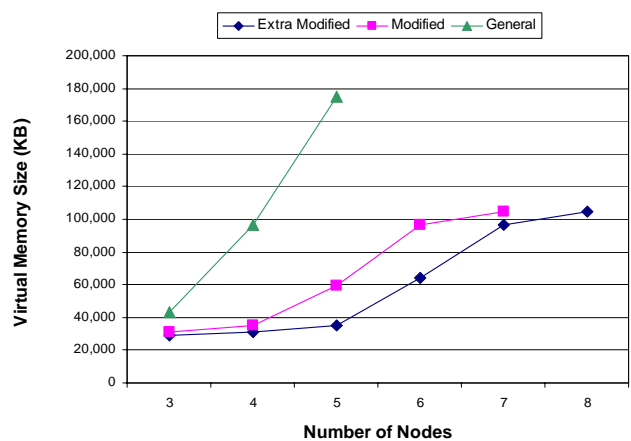

**Fig. 5.** Virtual memory usage in the model checking process for each case.

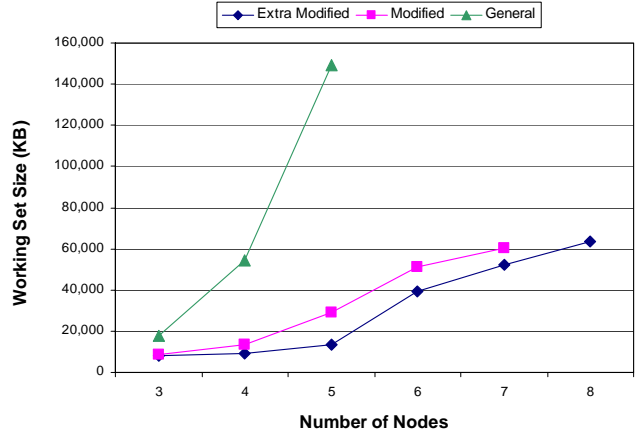

**Fig. 6.** Memory working set size in the model checking process for each case.

Figure 5 and figure 6 respectively show the amount of virtual memory and the working set size that the model checker uses in each case. It means that after building the model, how much memory the model checker will need to perform the property checking for the explained properties in the ring. As can be seen in both figures, for all the three cases like the previous figures the growths are dramatically fast for the general version. On the other hand, for the other two suggested versions, we have dramatically more decent growth behavior in the memory usage graphs. It means that the memory usage rate generally decreases when the number of nodes is increased in the ring. Although I did the model checking for up to 4 nodes for general case, these two figures contain the virtual memory usage and working set size of general version of the algorithm for 5 nodes. For doing CTL model checking in NuSMV, after running the ctlspec command, NuSMV will build the complete model graph and after that it starts to check the property. Hence, the memory usage for the ring with 5 nodes could be roughly estimated.

## 10 Conclusion

In general, performing SMV model checking for the Peterson's algorithm with more than four nodes does not seem to be feasible in reasonable amount of time. According to that, through this project, I have tried to simplify the algorithm as much as possible to reach to a new algorithm with the same functionality that allows us to do the normal model checking procedure with NuSMV. By using these simplifications on the way that the algorithm works, I could verify the three main expected properties of the algorithm in an eight node asynchronous ring for determining the leader. These three expected properties are: this leader should be eventually elected; in any state of the system, there should not be more than one leader in the ring; and the leader's virtual ID tag should be equal to the greatest initialized UID value of all the nodes in the ring. Using these main constraints on the behavior of the Peterson's algorithm can significantly decrease the amount of the CPU time, virtual memory size, working set size and the total number of the page faults. Considerable amount of decrease in the usage of main computer resources by the model checker enables us to target the higher number of nodes in our verification process. Finally by hiring these simplifications, I verified the three main desired properties of Peterson's algorithm for leader election in the asynchronous rings with three to eight nodes.

Since the symbolic model checking process mostly faces with the explosion in the set of states, the method which is used in this paper for decreasing the number of states in the model checking process can be leveraged for extending the boundary of the formal verification of the similar problems using the symbolic model checking.